\documentclass[11pt]{gh2013}

\usepackage{mathptmx}       
\usepackage{helvet}         
\usepackage{courier}        
\usepackage{makeidx}         
\usepackage{graphicx}        
\usepackage{multicol}        

\begin{document}

\title*{HII galaxies in 4D}
\titlerunning{Massive Young Star Clusters Near and Far}
\author{Eduardo Telles}
\institute{Observat\'orio Nacional, Rua Jos\'e Cristino, 77, Rio de Janeiro, RJ, 20921-400, Brazil, \email{etelles@on.br}}
%
%
\maketitle


\vskip -3.5 cm 

\abstract{ HII galaxies are clumpy and their gas kinematics
  can be mapped to show the global turbulent motions and the
  effect of massive star evolution. The distribution of their physical conditions
is homogenous and oxygen abundance is uniform. The presence of
  nebular HeII 4868 line seems to be higher in a low abundance galaxy,
  implying a harder ionization power probably due to stars in low
  metallicity. Innovative methods of data cube analysis, namely PCA
  tomography, seem promising in revealing additional information not
  detected with the standard methods. I review some of
  our own recent work on the 3D spectroscopy of HII galaxies.
}
\section{HII galaxies in 1D and 2D}

\begin{figure*}
\centering
  \includegraphics[width=  7.5 cm]{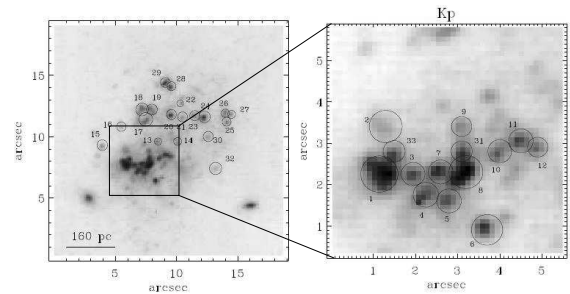}
  \caption{Mrk 36 (Haro 4) Gemini-niri K-band image from \cite{lag11}. This high spatial resolution image shows unprecedented details of its ``blobbiness''.}
  \label{fig:mrk36}
\end{figure*}

HII galaxies have been found by the extraordinary optical 1D
spectroscopic characteristics of their starburst region, strong line
emission on a faint blue continuum \cite{keh04}\cite{ter91}. Their
physical properties show their very low oxygen abundances and dust
content, and ionization by massive stars.  Attempts to derive the
properties of the stellar population and star formation histories of
their underlying host galaxies proved to be very difficult, but
indicated that most HII galaxies are not forming stars for the first
time \cite{wes04}.  The presence of an underlying galaxy in a sample
of HII galaxies has been detected in 2D optical imaging and surface
photometry, that showed that HII galaxies are at least middle aged
\cite{tel97}.  This result has been confirmed with high spatial
resolution near-IR imaging (fig~\ref{fig:mrk36}) \cite{lag11}.  The
surface photometry of these 2D near-IR imaging and previous
narrow-band H$\beta$ imaging with NTT/ESO \cite{lag07} have allowed us
to conclude that star forming knots in HII galaxies, as in more
luminous starbursts, are fragmented and consist of ensembles of star
clusters (also see posters by Adamo, by Santos, and by Torres
Campos). Typically, star clusters are numerous over hundreds of parsec,
with young ages ($<$ a few $10^7$ yrs) and no sequential trend in ages
across the galaxy. This alone rules out self-propagation as a dominant
mechanism for star formation in galactic scales.  Our view of the mode
of star formation in these regular, low luminosity, isolated HII
galaxies is that star cluster formation is triggerred by internal
dynamical processes.  In this picture, star formation is global,
simultaneous, and stochastic {\it within these time scales}, hence our
``pop-corn model'' of star formation \cite{tel10}. It is worth
mentioning that age dating young star clusters remains a
challenge. Single stellar population models are not red enough to
reproduce the observed optical/near-IR colors without some
unrealistic assumptions about extinction and metallicity.  The
K-band- or red- excess (as compared to SSP models) have been discussed
by many of us over recent years without a consensus.  Other
alternatives invoke nebular emission, dust emission, extended red
emission, red supergiants, and an intermediate age population (with or
without TP-AGBs). Another possibility is the effect, on the SED of the
young clusters, of stellar rotation and/or binarity at low metallicities to be
included in the models (see talk by Leitherer). We will keep tuned.

\section{HII galaxies in 3D}
\label{sec:3d}

\begin{figure*}
  \centering
  \includegraphics[width=  3.5 cm]{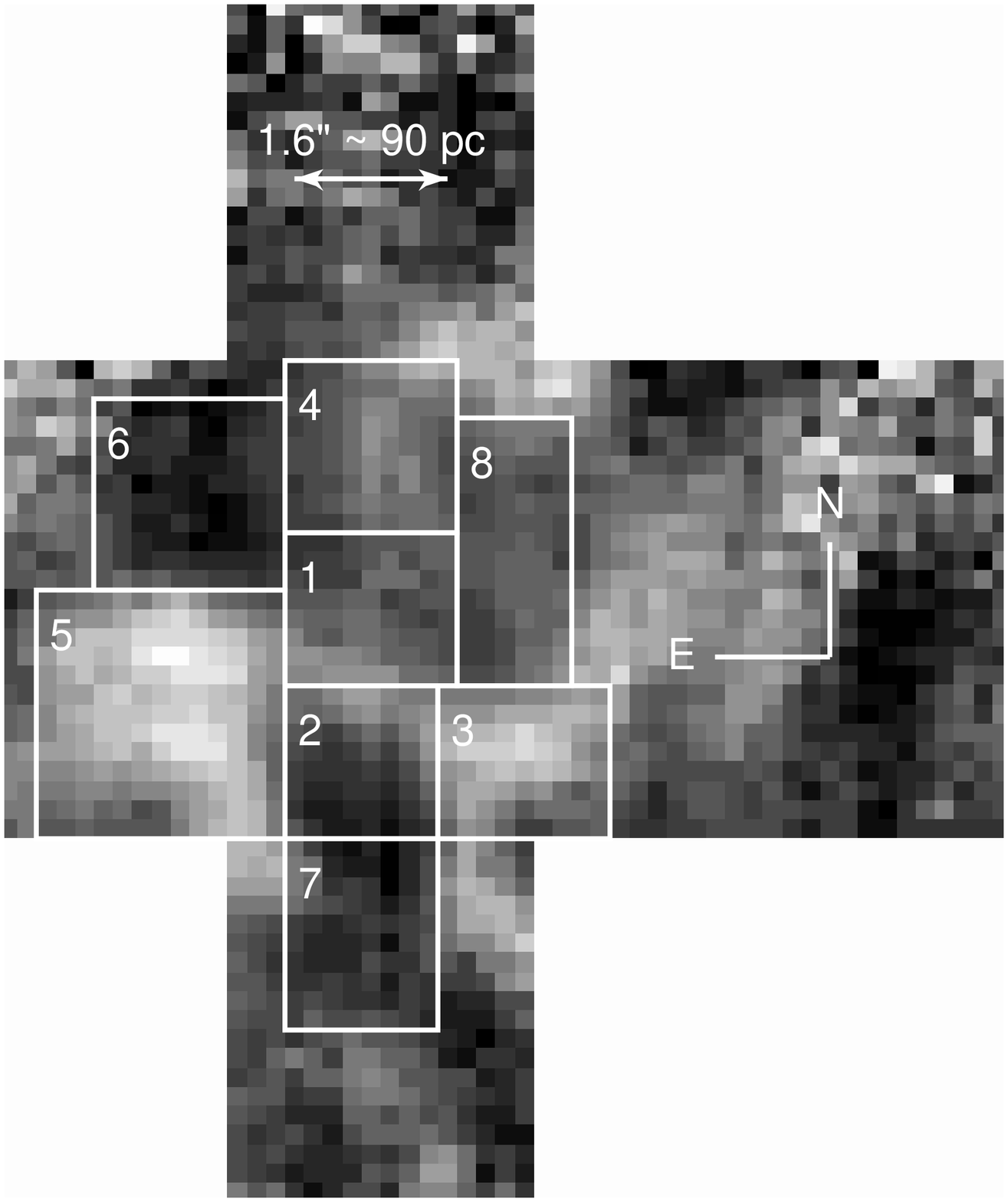}
  \includegraphics[width=  4.5 cm]{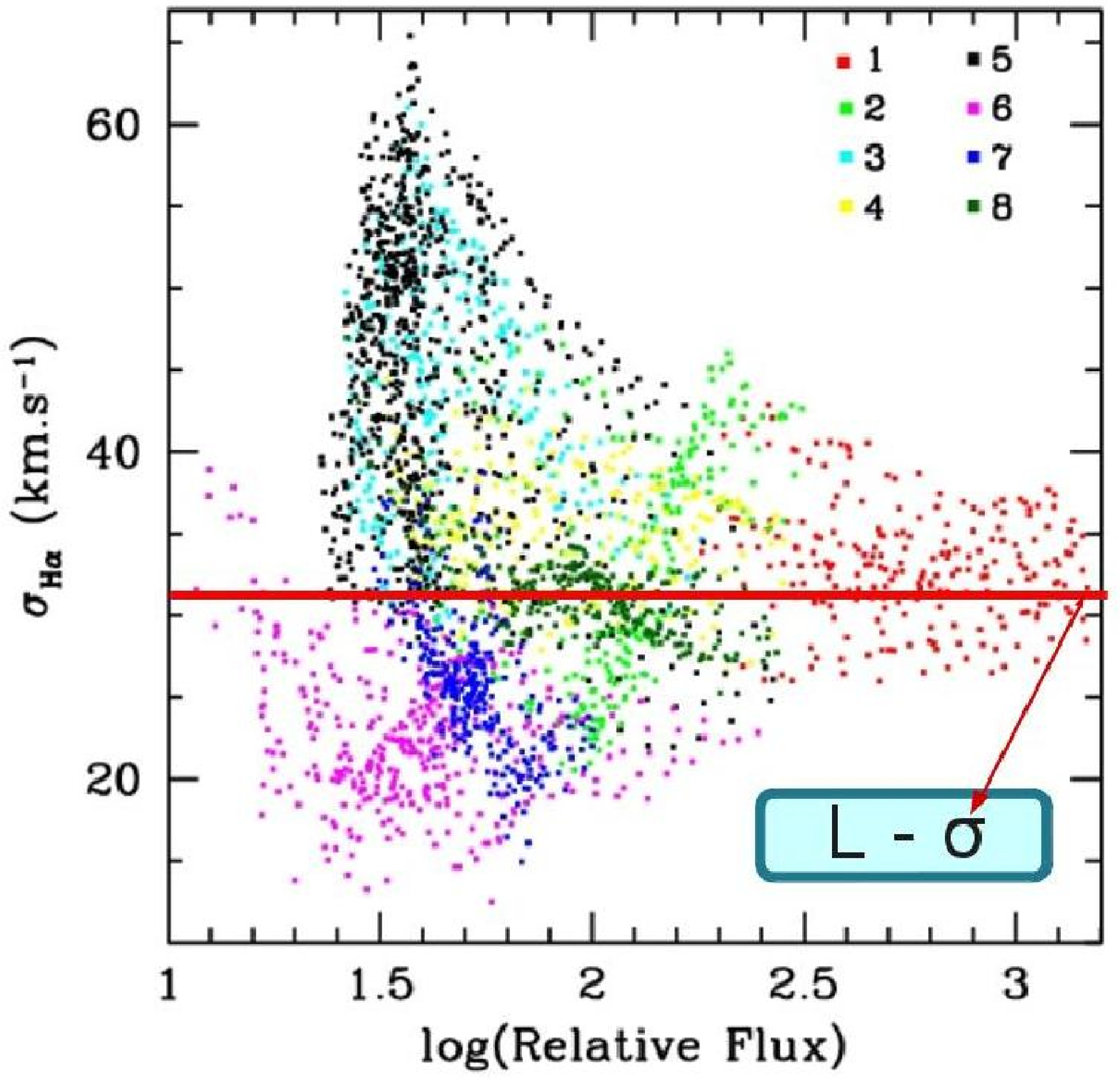}

  \caption{II Zw 40 Gemini GMOS \cite{bor09}. (left) GMOS mosaic with regions of interest. (right) I-$\sigma$ diagram. The line is the integrated line width.}
  \label{fig:zw40}
\end{figure*}

Much progress has been made in recent years with the advent of
integral field spectrographs (IFU).  Type II HII galaxies
\cite{tel97b} have been particular good targets because of their small
sizes and intense nebular emission.  There are now a good number of
objects observed with IFU (see an incomplete list compiled by
\cite{lag13}, and talk by B.James).  Our first motivation in the use of
the IFUs was to identify the origin of the supersonic motions of the
warm gas. Echelle spectroscopy gave us a hint that the
brightest star forming knot was responsible for the global kinematics
\cite{tt01}.  We chose to target the well known HII galaxy: II Zw 40
with Gemini GMOS-IFU \cite{bor09} and SINFONI \cite{van08}.  The
analysis of a set of kinematics diagnostic diagrams, such as the
intensity versus velocity dispersion (I-$\sigma$), intensity versus
radial velocity (I-V) and V-$\sigma$, for global and individual
analysis in sub-regions of the nebula  separates the main
line broadening mechanisms responsible for producing a smooth
supersonic integrated line profile
(fig~\ref{fig:zw40}). As can be seen, the brightest knot
dominates the light weighted integrated line width, deconvolved from
the local effects of stellar evolution represented by regions of low
intensity and higher $\sigma$. This means that when we measure the
line widths of any aperture including the kinematic core, we measure
the velocity dispersion of the underlying turbulent motions that we
interpret as being due to motions in the gravitational potential.
This is reassuring for the application of the L-$\sigma$ relation as
distance and mass indicators with cosmological interest (see
\cite{bor11} and talks by R. Terlevich,  by Plionis, and by Ch\'avez).

IFU data also allow the mapping of physical conditions, and the
connection between kinematics and star formation.  Our results
\cite{keh08} \cite{lag09}\cite{lag12} have shown that the physical
conditions are typically homogeneous and the oxygen abundances are
uniform. This can be envisaged if metals remain in the hot phase
ejected by SN explosions during the starburst, and then heavy element
droplets mix the abundance over the disk of the galaxy in certain
timescale. Alternatively, they have not undergone strong supergalactic
nor selective winds, and they have not accreted large amounts of
uncontaminated matter. They simply have been forming stars very
slowly. In this case young superstar clusters have little to do with
their metallicity. The main responsible producer is the underlying
stellar population (see \cite{tt12}, and references therein).
Ionization is mainly due to the UV ionizing flux from massive star,
and Wolf-Rayet stars may be responsible for the observed local
nitrogen enhancement, particularly in the dense nucleus of Mrk 996
\cite{tel14} (see talk by Bethan James). The origin of the observed
broad component in the emission lines in Mrk 996, as in other low
metalicity dwarfs is still under debate (see \cite{izo08}), and we
have used adaptive optics assisted near-IR IFU spectroscopy with
Gemini-NIFS to address this issue (in preparation).  A final topic in
our analysis of 3D spectroscopy is the origin of the narrow nebular
HeII $\lambda$4686 observed in some objects (see talks by Oey, and by
Jaskot). This seems to be an unresolved issue.  The emission seems to
be more intense in lower metallicity HII galaxies \cite{lag12}. In
this case, their source should be associated with O stars at low
metallicity, as opposed to other candidates such as WR stars,
primordial (zero-metallicity) stars, high-mass X-ray binaries, or
radiative shocks. However, not all lower metallicity HII galaxies show
narrow HeII $\lambda$4686 emission. Massive stars with rotation, or in
binaries, may produce significantly harder ionizing radiation that can
also be invoked to address this issue.  Again, we have to await for
newer stellar models.

\section{HII galaxies in 4D}
\label{sec:2}

\def\Zsun{\thinspace\hbox{$\hbox{Z}_{\odot}$}}

To fully exploit the wealth and complexity of the information that IFU
data provide, we need analysis techniques that are more sophisticated
than those commonly used in one-dimensional long-slit spectroscopy. In
\cite{tel14} we applied the technique of Principal Component Analysis
(PCA) tomography to extract spatial and spectral information from data
cubes of Mrk 996 in a statistical manner. A presentation of the
technique can be found in \cite{ste09}. What PCA tomography basically
does is to extract hidden information by transforming a large set of
correlated data, in our case the wavelength pixels, into a new set of
uncorrelated variables, ordered by their eigenvalues.  The PCA
tomography analysis that corroborates, in a completely independent
manner, the kinematic picture outlined for Mrk 996: an outflow from
the inner region, associated with winds from WR stars, superposed on
an underlying rotation pattern, affecting the motions of low-density
clouds. This recently developed statistical method for handling data
cubes has also resulted in the independent detection of the narrow
component of the [O III]$\lambda$4363 line, not seen previously in
integrated spectra. This detection allows a reliable and direct
measurement of the chemical abundances in the region outside the
nucleus. We obtain an oxygen abundance 12+log(O/H)=7.90$\pm$0.30
($\sim$0.2 \Zsun) for the low-density narrow-line region around the
nucleus of Mrk 996. Traditional techniques of 1D slit spectroscopy may
not be the proper tool to use to analyse the wealth of information of
IFU spectroscopy.  
This new PCA tomography technique is promising, and
hopefully one of many new methodologies expected to be devised in the
near future.

%



\begin{thebibliography}{99.}%

\bibitem{keh04} Kehrig, C., Telles, E., \& Cuisinier, F.\ 2004, AJ, 128, 1141 


\bibitem{ter91} Terlevich R., Melnick J., Masegosa J., Moles M., Copetti M.~V.~F., 1991, A\&AS, 91, 285 

\bibitem{wes04} Westera, P., Cuisinier, F., Telles, E., \& Kehrig, C.\ 2004, A\&A, 423, 133 

\bibitem{tel97} Telles, E., \& Terlevich, R.\ 1997, MNRAS, 286, 183 

\bibitem{lag11} Lagos, P., Telles, E., 
Nigoche-Netro, A., \& Carrasco, E. R.\ 2011, AJ, 142, 162 

\bibitem{lag07} Lagos, P., Telles, E., \& Melnick, J.\ 2007, A\&A, 476, 89 

\bibitem{tel10} Telles, E.\ 2010, ASPC, 423, 65 

\bibitem{tel97b} Telles E., Melnick J., Terlevich R., 1997, MNRAS, 288, 78 

\bibitem{lag13} Lagos P., Papaderos P., 2013, Advances in Astronomy, 1

\bibitem{tt01} Telles E., Mu{\~n}oz-Tu{\~n}{\'o}n C., Tenorio-Tagle G., 2001, ApJ, 548, 671 

\bibitem{bor09} Bordalo V., Plana H., Telles E., 2009, ApJ, 696, 1668 

\bibitem{van08} Vanzi L., Cresci G., Telles E., Melnick J., 2008, A\&A, 486, 393 
\bibitem{bor11} Bordalo V., Telles E., 2011, ApJ, 735, 52 

\bibitem{keh08} Kehrig C., V{\'{\i}}lchez J.~M., S{\'a}nchez S.~F., Telles E. et al., 2008, A\&A, 477, 813 

\bibitem{lag09} Lagos P., Telles E., Mu{\~n}oz-Tu{\~n}{\'o}n C.,
  Carrasco E.~R., Cuisinier F., Tenorio-Tagle G., 2009, AJ, 137, 5068


\bibitem{lag12} Lagos P., Telles E., Nigoche Netro A., Carrasco E.~R., 2012, MNRAS, 427, 740 



\bibitem{tt12} Tenorio-Tagle G., 2012, Dwarf Galaxies: Keys to Galaxy
  Formation and Evolution,
  Astro\-phys.\nobreak\ Sp.\nobreak\ Sc.\nobreak\ Proceedings, ISBN
  978-3-642-22017-3. Springer-Verlag, 2012, p. 253


\bibitem{tel14} Telles E., Thuan T.~X., Izotov Y.~I., Carrasco E.~R., 2014, A\&A, 561, A64 

\bibitem{izo08} Izotov Y.I., Thuan T.~X., ApJ, 687, 133

\bibitem{ste09} Steiner, J.E., Menezes, R.B., Ricci, T.V. \& Oliveira, A.S. 2009, MNRAS, 395, 64









\end{thebibliography}
\end{document}